\documentclass[a4paper]{JHEP3}
\usepackage{epsfig,bbm}

\title{Resummation of rapidity logarithms in $B$ meson wave functions}

\author{Hsiang-nan Li$^{a,b,c}$, Yue-Long Shen$^{d}$,  Yu-Ming Wang$^{e}$
\footnote{Address after September 30, 2012: Physik Department T31, James-Franck-Stra${\ss}$e,
Technische Universit\"{a}t M\"{u}nchen, D-85748 Garching, Germany.}

{\it \small $^a$Institute of Physics, Academia Sinica,
Taipei, Taiwan 115, Republic of China }
\\
{\it \small $^b$Department
of Physics, National Cheng-Kung University, Tainan, Taiwan 701,
Republic of China}
\\
{\it \small $^c$Department of Physics, National
Tsing-Hua University, Hsinchu, Taiwan 300, Republic of China}
\\
{\it \small $^d$College of Information Science and Engineering,
Ocean University of China, Qingdao, Shandong 266100, P.R. China}
\\
{\it \small $^e$ Theoretische Elementarteilchenphysik,
Naturwissenschaftlich Techn. Fakult\"at \\ Universi\"at Siegen,
57068 Siegen, Germany}

}

\abstract{We construct an evolution equation for the $B$ meson wave functions
in the $k_T$ factorization theorem, whose solutions sum the double
logarithms associated with the light-cone singularities, namely,
the rapidity logarithms. The derivation is subtler than that of
the Sudakov resummation for an energetic light hadron, due to the
involvement of the effective heavy-quark field. The renormalization-group
evolution in the factorization scale needs to be included in order
to derive an ultraviolet-finite and scale-invariant kernel for resumming the
rapidity logarithms. It is observed that this kernel is similar to that
of the joint resummation for QCD processes in extreme kinematic regions,
which combines the threshold and $k_T$ resummations. We show that the
resummation effect maintains the normalization of the $B$ meson wave
functions, and strengths their convergent behavior at small spectator
momentum. The resummation improved $B$ meson wave functions are then
employed in the leading-order analysis of the $B\to\pi$ transition form
factors, which lead to approximately 25 \% deduction in the large recoil region.}


\begin{document}


\section{INTRODUCTION}

The $B$ meson distribution amplitudes $\phi_B^\pm(k^+)$ are
essential inputs for a perturbative analysis of exclusive $B$ meson
decays based on the collinear factorization theorem
\cite{BL,ER,CZS,CZ,L1}, where $k^+$ is the momentum carried by the
light spectator quark. Their properties have
been investigated intensively: models of
$\phi_B^\pm$ with an exponential tail in the large $k^+$ region were
proposed in \cite{GN}. Neglecting three-parton distribution
amplitudes in a study based on equations of motion
\cite{PB1,Braun:1990iv}, $\phi_B^\pm$ were found to contain
step functions with a sharp drop at large $k^+$ \cite{Kawamura:2001jm}.
The determination of the $B$ meson distribution amplitudes from relevant data
was discussed in \cite{ML99,CL05}.
The asymptotic behavior of $\phi_B^+$ was extracted from a
renormalization-group (RG) evolution equation derived in the
framework of the collinear factorization theorem, which
decreases slower than $1/k^+$ \cite{Neu03}. That is, the $B$ meson
distribution amplitude is not normalizable, after taking into
account the RG evolution effect. This feature
was confirmed in a QCD sum rule analysis \cite{BIK}, which
includes next-to-leading-order (NLO) perturbative corrections.
It has been argued that a non-normalizable $B$ meson distribution
amplitude does not cause trouble in calculations of decay amplitudes,
if the first inverse moment $\int dk^+\phi_B^+(k^+)/k^+$ was involved \cite{BBNS,PK}.
However, the non-normalizability does introduce difficulty in defining the
$B$ meson decay constant $f_B$ via the integration of the $B$ meson distribution
amplitude \cite{Li:2004ja}.

It has been known that collinear factorization formulas for various
exclusive $B$ meson decay amplitudes, such as factorizable contributions
and nonfactorizable contributions at higher powers, suffer end-point
singularities \cite{SHB}. These singularities imply
that the $k_T$ factorization theorem
\cite{CCH,CE,LRS,BS,LS,NL2} is a more appropriate theoretical
framework for exclusive $B$ meson decays \cite{Monr}. Retaining parton
transverse momenta $k_T$ \cite{HS}, as postulated in the perturbative
QCD (PQCD) approach \cite{LY1,CL,KLS,LUY} based on the $k_T$
factorization theorem, the end-point singularities
disappear \cite{TLS}, and resultant predictions are in
agreement with most of data \cite{KS02,Li07,Ali:2007ff}. It was
then pointed out that the non-normalizability of
$\phi_B^+$ is also a consequence of the collinear
factorization theorem \cite{Li:2004ja}. Reanalyzing
the RG evolution effect on the (unintegrated) $B$ meson wave function
in the $k_T$ factorization theorem \cite{Li:2004ja},
it was found that the ultraviolet behavior of its evolution
kernel is tamped. As a result, the RG evolution maintains the
normalization of the $B$ meson wave function.

As elaborated in \cite{Co03}, a $k_T$-dependent hadron wave function
contains additional infrared divergences from the region with a loop momentum
parallel to the Wilson line on the light cone. These light-cone
singularities, cancelling each other in the collinear factorization theorem,
appear in the $k_T$ factorization theorem. To regularize the
light-cone singularities, we have rotated the Wilson line from the
light cone to an arbitrary direction $n$ with $n^2\not=0$ \cite{CS91,Co03}.
The higher-order wave function then depends on $n^2$ through the scale
$\zeta_P^2 \equiv 4(P \cdot n)^2/n^2$, where $P$ denotes the hadron
momentum. The variation of $\zeta_P^2$ introduces a factorization-scheme
dependence of a hadron wave function. The evaluation of the
NLO effective diagrams for the $B$ meson and pion wave
functions indicates the existence of the double logarithm $\ln^2\zeta_P^2$
\cite{LSW12} and the single logarithm $\ln\zeta_P^2$ \cite{LSW11}, 
respectively\footnote{This logarithm corresponds
to the scheme-dependent logarithm $\ln {\nu^{\prime}}^2$ in \cite{Li:2004ja},
where a different definition for the $B$ meson wave function was adopted.}.
These logarithms, with the same origin as the rapidity logarithms discussed in
\cite{CS91,Co92,Chiu:2012ir,FZ12}, need to be resummed as $\ln\zeta_P^2$ becomes large.
It is expected that the resummation of the rapidity logarithms in the $B$
meson wave functions will reduce the scheme dependence.

In this paper we shall construct an evolution equation in the ratio
$\zeta^2\equiv \zeta_P^2/m_B^2=4(v \cdot n)^2/n^2$, $m_B$ ($v$) being the $B$
meson mass (velocity), that resums the rapidity logarithms in the $B$
meson wave functions. The idea is similar to that of the soft-collinear
resummation for a light hadron wave function \cite{CS91,Altinoluk:2012fb}.
The difference arises from the effective heavy-quark field
involved in the definition of the former, which modifies
the ultraviolet behavior of the evolution kernel. It will be shown
that the RG evolution in the factorization scale $\mu_f$, whose anomalous
dimension also depends on $\zeta^2$, has to be taken into account in
order to derive an ultraviolet finite kernel. That is,
the evolutions in both $\zeta^2$ and $\mu_f$ must be considered
simultaneously for a consistent and complete treatment of
the logarithmic corrections to the $B$ meson wave functions,
an observation in agreement with that in \cite{Chiu:2012ir}.
The solutions to the evolution equation contain the resummation of
the rapidity logarithms, and their limits as $\zeta^2\to\infty$
exist. It will be demonstrated that
the wave function $\phi_B^-$, which takes a finite value at $k^+=0$ before
the resummation, vanishes after the resummation.
The wave function $\phi_B^+$, which diminishes at $k^+=0$ before
the resummation, approaches zero faster after the resummation.
Namely, the effect from resumming the rapidity logarithms
suppresses the behavior of the $B$ meson wave functions near the
end point $k^+=0$.

In Sec.~II we construct the evolution equation that resums the rapidity logarithms
in the $B$ meson wave functions.
The equation is then solved in the Mellin space in Sec.~III, with the models of
the $B$ meson wave functions proposed in \cite{Kawamura:2001jm}
as the initial condition of the evolution. As performing the inverse Mellin
transformation of the above solutions, the extrapolation
of the running coupling constant down to the low energy region is specified to
avoid the Landau pole. It is observed that the two $B$ meson wave functions,
whose $k^+$ dependencies differ dramatically before the resummation, become similar:
both diminish faster than $k^+$ at $k^+=0$. The $B\to\pi$ transition form factors are then calculated
for the $B$ meson wave functions before and after the resummation. It is found that
the resummation effect decreases the form factors by approximately 25 \% at large recoil, which is
attributed to the suppression of the $B$ meson wave functions near the end point. We
summarize our findings in Sec.~IV, and collect the explicit expressions for the
solutions of the evolution equation in Appendix A.

\section{EVOLUTION EQUATION}

The $B$ meson wave functions $\Phi_B^\pm$ constructed in the $k_T$ factorization
theorem \cite{CCH,CE,LRS,BS,LS,HS},
\begin{eqnarray}
& &\langle 0|{\bar q}(y) W_y(n)^{\dag}I_{n;y,0}W_0(n) \Gamma h(0)|{\bar B}(v)\rangle \nonumber \\
&& = -{i f_B m_B \over 4} {\rm Tr} \left \{ {1 +\slash \! \! \!   v \over 2}
\left [ 2 \, \Phi_{B}^{+}(t,y^2) + { \Phi_{B}^{-}(t,y^2)
-\Phi_{B}^{+}(t,y^2)   \over t }  \slash \! \! \!   y \right ] \gamma_5 \, \Gamma\right \},
\label{de1}
\end{eqnarray}
describe the distributions of the light parton in both the longitudinal direction
denoted by $t=v \cdot y$ and the transverse direction denoted by $y^2$.
In the above definition $y=(0,y^-,{\bf y}_T)$ is the coordinate of the anti-quark
field $\bar q$, $h$ is the rescaled $b$ quark field
characterized by the $B$ meson velocity $v$, and
$\Gamma$ represents a Dirac matrix. The Wilson line operator $W_y(n)$ is written as
\begin{eqnarray}
\label{eq:WL.def}
W_y(n) = {\cal P} \exp\left[-ig \int_0^\infty d\lambda
n\cdot A(y+\lambda n)\right],
\end{eqnarray}
where $\cal P$ means the path-ordered integration, and $g$ is the strong coupling
constant. The vertical link $I_{n;y,0}$ at infinity does not contribute in the
covariant gauge \cite{CS08}. Note that a non-light-like vector $n$
has been substituted for the null vector $n_-=(0,1,{\bf 0}_T)$, so that the
light-cone divergences associated with the Wilson lines are regularized by $n^2\not=0$.

According to \cite{LSW12}, the convolution of the NLO $B$ meson wave
functions with the leading-order (LO) hard kernel produces both the single
and double logarithms of $\ln\zeta^2$, which become large as
$(v\cdot n)^2\gg n^2$. The latter is generated by the gluon
exchange between the rescaled $b$ quark and the Wilson line ending
at the spectator coordinate $y$ \cite{LSW12}.
The double logarithm arises from the overlap of the collinear enhancement from a loop
momentum $l$ collimated to $n$ and the soft enhancement from small $l$.
To resum the above rapidity logarithms, we follow the strategy
developed in \cite{Li:1996gi}: we vary the velocity $v$
under the constraint $v^2=1$, which respects the equation of motion for
the rescaled $b$ quark. Since the collinear dynamics is independent of $v$, the
variational effect on the $B$ meson wave function does not involve the collinear
dynamics, and is factorizable. An evolution equation of the $B$ meson
wave function in $v$, or equivalently in $\zeta^2$, is then derived,
whose solution resums the rapidity logarithms. As claimed in
the Introduction, the resummation effect exists in the $\zeta^2\to\infty$
limit, where the scheme dependence of the $B$ meson wave functions
is shown to disappear. The condition $v^2=1$ implies that
the two components $v^+$ and $v^-$ are not independent,
and that the derivative $dv^2/dv^+=0$ gives $dv^-/dv^+=-v^-/v^+$.
Therefore, the derivative of the wave function $\Phi_B$ should
be understood as
\begin{eqnarray}
v^+\frac{d}{dv^+}\Phi_B=\left(v^+\frac{\partial}{\partial
v^+}-v^-\frac{\partial}{\partial v^-}\right)\Phi_B\equiv
\epsilon_{\alpha\beta}v^\alpha\frac{\partial}{\partial v_\beta}\Phi_B,
\end{eqnarray}
with the anti-symmetric tensor $\epsilon_{\alpha\beta}$,
$\epsilon_{+-}=-\epsilon_{-+}=1$.

We have to differentiate the bare wave function $\Phi_B^{(b)}$ and the
renormalized wave function $\Phi_B$
here, because they are defined in the heavy-quark effective theory.
This differentiation is not necessary in the pion case as performing the
Sudakov resummation, which is
defined in full QCD. The difference becomes manifest as comparing
their NLO expressions: the former contains a $\zeta^2$-dependent
ultraviolet pole from the vertex correction formed by the rescaled $b$
quark line and the Wilson line \cite{LSW12}
\begin{eqnarray}
- \frac{\alpha_s C_F}{4 \pi}\ln \zeta^2
\left(\frac{1}{\epsilon}+\ln\frac{4\pi\mu_{\rm f}^2 }{m_g ^2
e^{\gamma_E}}\right), \label{5d}
\end{eqnarray}
$m_g$ being an infrared regulator and $\gamma_E$ being the Euler constant,
but the latter does not. The variation of $v$ is equivalent to
that of $\zeta^2$,
\begin{eqnarray}
\frac{v\cdot n}{2\epsilon_{\alpha\beta}v^\alpha
n^\beta}v^+\frac{d}{dv^+}\Phi_B^{(b)}(x,k_T,\zeta^2,\mu_f)= \zeta^2
\frac{d}{d\zeta^2}\Phi_B^{(b)}(x,k_T,\zeta^2,\mu_f),\label{deri}
\end{eqnarray}
with the momentum fraction $x\equiv k^+/P^+$ of the spectator.
Inserting $\Phi_B^{(b)}=Z_\Phi\Phi_B$
into Eq.~(\ref{deri}), we have
\begin{eqnarray}
\frac{1}{Z_\Phi}\zeta^2\frac{d}{d\zeta^2}\Phi_B^{(b)}(x,k_T,\zeta^2,\mu_f)&=&
\frac{1}{Z_\Phi}\left(\zeta^2\frac{d}{d\zeta^2}Z_\Phi\right)\Phi_B(x,k_T,\zeta^2,\mu_f)
\nonumber \\
&& +\zeta^2\frac{d}{d\zeta^2}\Phi_B(x,k_T,\zeta^2,\mu_f), \label{deri1}
\end{eqnarray}
where $Z_\Phi$ is the $\zeta^2$-dependent renormalization constant of the $B$ meson
wave function resulting from Eq.~(\ref{5d}).

The $v$ dependence in $\Phi_B^{(b)}$ is
introduced through the eikonal line associated with the rescaled $b$ quark.
Hence, the derivative with respect to $v^+$ applies to the Feynman
rule for the rescaled $b$ quark propagator
\begin{eqnarray}
\frac{v\cdot n}{2\epsilon_{\alpha\beta}v^\alpha
n^\beta}v^+\frac{d}{dv^+}\frac{v^\mu}{v\cdot l}=\frac{\hat
v^\mu}{v\cdot l},
\end{eqnarray}
leading to the special vertex
\begin{eqnarray}
\hat v^\mu \equiv \frac{v\cdot n}{2\epsilon_{\alpha\beta}v^\alpha
n^\beta}\epsilon_{\rho \lambda} v^\rho \left(g^{\mu
\lambda}-\frac{v^\mu l^\lambda}{v\cdot l}\right).
\end{eqnarray}
It is easy to see that the contributions from the two terms in the special
vertex $\hat v^\mu$, as contracted to a vertex in $\Phi_B^{(b)}$,
cancel each other, when $l$ is collimated to $n$. That is,
$\hat v^\mu$ suppresses the dominant collinear dynamics associated with the
Wilson lines. Therefore, the variational effect involves only the soft
dynamics, and we can eikonalize the attachments of the differentiated
gluon emitted by the special vertex to all internal lines in the wave
function. Following the reasoning in \cite{Li:1996gi},
the special vertex must appear at the outer most end
of the eikonal line at leading-logarithm accuracy. Applying the Ward identity
to the sum over all attachments of the differentiated gluon, we factorize
the differentiated gluon out of $\Phi_B^{(b)}$.

\begin{figure}[ht]
\begin{center}
\hspace{-1 cm}
\includegraphics[scale=0.6]{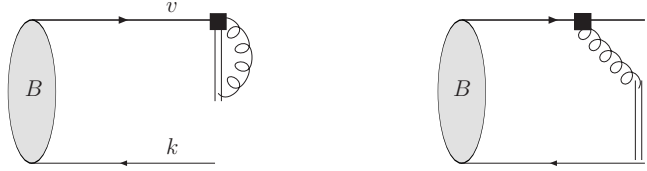}
\caption{LO soft kernel for the $\zeta^2$-evolution equation, where the
square at the end of the rescaled $b$-quark line represents the special vertex.}
\label{fig1}
\end{center}
\end{figure}

The derivative of $\Phi_B^{(b)}$ is then written as the convolution
\begin{eqnarray}
\frac{1}{Z_\Phi}\zeta^2 \frac{d}{d\zeta^2}\Phi_B^{(b)}(x,k_T,\zeta^2,\mu_f)
&=&\frac{1}{Z_\Phi}K^{(b,1)}\otimes \Phi_B^{(b)}(x,k_T,\zeta^2,\mu_f) \nonumber\\
&=& K^{(b,1)} \otimes \Phi_B(x,k_T,\zeta^2,\mu_f),\label{derb}
\label{res}
\end{eqnarray}
where the one-loop kernel $K^{(b,1)}$ collects the soft divergences generated by the
differentiated gluons in Fig.~\ref{fig1}. The two diagrams in Fig.~\ref{fig1} give
$K^{(b,1)}=K_1^{(b,1)}+K_2^{(1)}$, with
\begin{eqnarray}
K_1^{(b,1)}&=&-ig^2C_F\int \frac{d^4l}{(2\pi)^4}\frac{\hat v\cdot n}{(v\cdot l+i\epsilon)
(l^2+i\epsilon) (n\cdot l+i\epsilon)},\\
K_2^{(1)}\otimes \Phi_B&=&ig^2C_F\int \frac{d^4l}{(2\pi)^4}
\frac{\hat v\cdot n}{(v\cdot l+i\epsilon)(l^2+i\epsilon) (n\cdot l+i\epsilon)}
\nonumber \\
&& \times  \Phi_B(x+l^+/P^+,k_T+l_T,\zeta^2,\mu_f).\label{k2}
\end{eqnarray}
Note that the gluon exchange between the rescaled $b$ quark and the spectator does
not introduce a $\zeta^2$ dependence \cite{LSW12}.
It is obvious that the soft poles from $l\to 0$ cancel between
$K_1^{(b,1)}$ and $K_2^{(1)}$, such that the evolution kernel
is infrared finite. However, compared to the conventional resummation
formalism for an energetic light hadron, Eq.~(\ref{res}) does not
contain a hard kernel $G$. Once the heavy-quark expansion is implemented,
the hard dynamics has been integrated out. The NLO corrections to the $B$ meson 
wave function from gluons radiated by the spectator quark do not produce the 
double logarithm $\ln^2 \zeta$ either \cite{LSW12}. Then the
ultraviolet divergence in $K_1^{(b,1)}$ seems to render the evolution kernel
ill-defined. As stressed in the Introduction, this problem can be resolved
by taking into account the $\zeta^2$-dependent RG evolution.

Substituting Eq.~(\ref{derb}) into Eq.~(\ref{deri1}), we arrive at
\begin{eqnarray}
\zeta^2\frac{d}{d\zeta^2}\Phi_B(x,k_T,\zeta^2,\mu_f) &=&
K^{(b,1)}\otimes \Phi_B(x,k_T,\zeta^2,\mu_f) \nonumber \\
&& -
\frac{1}{Z_\Phi}\left(\zeta^2\frac{d}{d\zeta^2}Z_\Phi\right)\Phi_B(x,k_T,\zeta^2,\mu_f),
\label{RGE:momenrum}
\end{eqnarray}
in which the second term plays the role of the counterterm for
$K_1^{(b,1)}$. The cancellation of the ultraviolet
poles is explicitly shown by computing $K_1^{(b,1)}$ and adopting $Z_\Phi$
in the $\overline{\rm MS}$ scheme \cite{Li:2004ja}
\begin{eqnarray}
K_1^{(b,1)}&=& -{\alpha_s C_F \over 4 \pi} \, \Gamma(\epsilon) \left({
4 \pi \mu_f^2 \over \lambda^2}\right)^{\epsilon} \, \left( { v \cdot n
 \over \epsilon_{\alpha\beta}v^\alpha n^\beta  } \right)^2,\label{kb1}\\
\delta K^{(1)}&\equiv&\frac{1}{Z_\Phi}\zeta^2\frac{d}{d\zeta^2}Z_\Phi  \nonumber\\
&=&-\frac{\alpha_s C_F}{4\pi} \,
\left [ {1 \over \epsilon} -\gamma_E + \ln (4 \pi) \right ] \,
\left( { v \cdot n  \over \epsilon_{\alpha\beta}v^\alpha n^\beta  }
\right)^2,\label{delk}
\end{eqnarray}
where $\lambda$ is an infrared regulator and the term
proportional to $n^2$ has been neglected relative to $v\cdot n$. The renormalization scale
in $K_1^{(b,1)}$ has been set to the factorization scale $\mu_f$, at which
the $B$ meson wave function is defined.

We apply the Mellin and Fourier transformations to the $B$ meson wave function
\begin{eqnarray}
\tilde \Phi_B(N,b,\zeta^2,\mu_f)=\int_0^1 dx (1-x)^{N-1}\int \frac{d^2 k_T}{(2\pi)^2}
\exp(i{\bf k}_T\cdot {\bf b})\Phi_B(x,k_T,\zeta^2,\mu_f),
\end{eqnarray}
under which the convolution in Eq.~(\ref{k2}) reduces to
\begin{eqnarray}
\int_0^1 dx (1-x)^{N-1}\int \frac{d^2 k_T}{(2\pi)^2}
\exp(i{\bf k}_T\cdot {\bf b})K_2^{(1)}\otimes \Phi_B = \tilde K_2^{(1)}(N,b,\zeta^2)
\tilde \Phi_B(N,b,\zeta^2,\mu_f), \hspace{0.5 cm}
\end{eqnarray}
with the soft kernel
\begin{eqnarray}
\tilde K_2^{(1)}(N,b,\zeta^2)&=&ig^2C_F\int
\frac{d^4l}{(2\pi)^4}\left(1+\frac{l^+}{P^+}\right)^{N-1}
\exp(-i{\bf l}_T\cdot {\bf b}) \frac{\hat v\cdot n}{(v\cdot
l+i\epsilon)(l^2+i\epsilon) (n\cdot l+i\epsilon)} \nonumber\\
&=&{\alpha_s C_F \over 2 \pi} \, \left ( { v \cdot n  \over
\epsilon_{\alpha\beta}v^\alpha n^\beta  } \right )^2 \left [  K_0
\left(\lambda b \right) -  K_0\left(\sqrt{\zeta^2}
 \frac{m_B b}{N} \right) \right ],
\end{eqnarray}
in the large $\zeta^2$ limit, $K_0$ being the zero-order modified Bessel function of second kind.
To simplify the integration, we have taken
the equivalent limit $v^{+} \to \infty$ with $n^{+}=n^{-}$.

Equation~(\ref{RGE:momenrum}) is rewritten, in the momentum ($N$) and
coordinate ($b$) spaces, as
\begin{eqnarray}
\zeta^2\frac{d}{d\zeta^2}\tilde{\Phi}_B(N,b,\zeta^2,\mu_f)=
\tilde{K}^{(1)}(N,b,\zeta^2,\mu_f) \, \tilde{\Phi}_B(N,b,\zeta^2,\mu_f),
\label{RGE:renom scheme}
\end{eqnarray}
with the renormalized soft kernel
\begin{eqnarray}
\tilde{K}^{(1)}(N,b,\zeta^2,\mu_f)&=&K^{(1)}_1(\mu_f)+\tilde
K_2^{(1)}(N,b,\zeta^2) \nonumber\\
&=&  - {\alpha_s C_F \over 2 \pi} \, \left [ \ln \frac{\mu_f
b}{2} +\gamma_E +  K_0\left(\sqrt{\zeta^2}
 \frac{m_B b}{N} \right) \right ],\label{K0}
\end{eqnarray}
where $K^{(1)}_1$ is defined by $K^{(1)}_1\equiv K^{(b,1)}_1-\delta K^{(1)}$,
and the infrared regulator $\lambda$ has disappeared in the
sum $K^{(1)}_1+\tilde K_2^{(1)}$.
Equation~(\ref{K0}) approaches $\ln b$ in the limit
$\sqrt{\zeta^2} m_B b\gg N$, and $\ln N$ in the limit $N\gg \sqrt{\zeta^2} m_B b$. Namely,
it involves the logarithms similar to those handled in the
joint resummation for QCD processes in extreme kinematic regions
\cite{Li99,LSV00,DFK11}, which combines the threshold
and $k_T$ resummations. We mention that the $\zeta^2$-evolution equations
derived in \cite{Feng:2009rp} differ from Eq.~(\ref{RGE:renom scheme}):
the hard kernel $G$ in their equation should not exist as explained before, once
the rescaled $b$-quark field is adopted; they did not resum $\ln N$,
so their equations are simpler;
their equations for $\Phi_B^+$ and $\Phi_B^-$ are different due to the incomplete $B$-meson
light-cone projector adopted in their calculation, but we have confirmed
that $\Phi_B^\pm$ obey the same equations; at last, they did not solve their
equations, so the behavior of the solutions is not clear.

The relation $\Phi_B^{(b)}=Z_\Phi\Phi_B$ also leads to the RG equation
\begin{eqnarray}
\mu_f\frac{d}{d
\mu_f}\Phi_B=-\frac{1}{Z_\Phi}\mu_f\frac{dZ_\Phi}{d
\mu_f}\Phi_B \equiv - \hat{\gamma}_B \Phi_B,\label{rgb}
\end{eqnarray}
with the $\zeta^2$-dependent anomalous dimension
\begin{eqnarray}
\hat{\gamma}_B= {\alpha_s C_F \over 2 \pi}  \left ( \ln\zeta^2 -2 \right ).
\end{eqnarray}
Solving Eq.~(\ref{rgb}) for arbitrary $\zeta^2$, we derive the RG evolution
\begin{eqnarray}
\tilde{\Phi}_B(N,b,\zeta^2,\mu_f)= \exp \left [ -\int^{\mu_f}_{\mu_0} \frac{d\mu}{\mu}
 \, {\alpha_s(\mu) \over 2 \pi}C_F \left ( \ln \zeta^2 -2   \right )    \right
 ] \, \tilde{\Phi}_B(N,b,\zeta^2,\mu_0),
 \label{scale solution}
\end{eqnarray}
with the initial scale $\mu_0$.
Note that the RG evolution in the collinear factorization theorem
drives the $B$ meson distribution amplitude to converge slowly at large
spectator momentum \cite{Neu03}.
Obviously, the RG evolution in the $k_T$ factorization
behaves normally as indicated in Eq.~(\ref{scale solution}).

Since the ultraviolet divergence and the light-cone divergence are from
different kinematical region, the scheme and scale
evolutions are commutable \cite{Chiu:2012ir}. The commutativity of
the derivatives with respect to $\zeta^2$  and $\mu_f$,
\begin{eqnarray}
\mu_f\frac{d}{d \mu_f}\zeta^2\frac{d}{d\zeta^2}\Phi_B=
\zeta^2\frac{d}{d \zeta^2}\mu_f\frac{d}{d \mu_f}\Phi_B,\label{double}
\end{eqnarray}
leads to the condition
\begin{eqnarray}
\mu_f\frac{d}{d \mu_f}K^{(b,1)}=0.\label{nec}
\end{eqnarray}
This is exactly the condition necessary for deriving the RG equation
of the soft kernel,
\begin{eqnarray}
\mu_f\frac{d}{d \mu_f} K^{(1)}=-\lambda_K,
\end{eqnarray}
with the anomalous dimension
\begin{eqnarray}
\lambda_K\equiv\mu_f\frac{d}{d \mu_f}\delta K^{(1)} = {\alpha_s C_F \over 2 \pi}.
\end{eqnarray}
The RG-improved soft kernel is then given by
\begin{eqnarray}
{\cal K}^{(1)}(N,b,\zeta^2,\mu_f)=
\tilde{K}^{(1)}(N,b,\zeta^2,\mu_c)-\int^{\mu_f}_{\mu_c}
\frac{d\mu}{\mu}\lambda_K(\alpha_s(\mu)) \, \theta(\mu_f-\mu_c),
\label{improved scheme kernel}
\end{eqnarray}
where the characteristic scale $\mu_c= a \, \sqrt{\zeta^2}\,m_B/N$,
$a$ being an order-unity constant, is chosen to remove large logarithms in the initial
condition $\tilde K^{(1)}(N,b,\zeta^2,\mu_c)$. As demonstrated later,
the suppression from the
resummation on the $B$ meson wave functions near $x=0$ does not depend on $a$.
The step function is introduced to terminate the evolution as $\mu_f <\mu_c$.

Solving Eq.~(\ref{RGE:renom scheme}) with the RG improved kernel in
Eq.~(\ref{improved scheme kernel}), one finds
\begin{eqnarray}
\tilde{\Phi}_B(N,b,\zeta^2,\mu_f)= \exp \left [
\int_{\zeta_0^2}^{\zeta^2} { d \tilde{\zeta}^2 \over
\tilde{\zeta}^2 } {\cal K}^{(1)}(N,b,\tilde{\zeta}^2,\mu_f)
\right ] \tilde{\Phi}_B(N,b,\zeta_0^2,\mu_f), \label{scheme
solution}
\end{eqnarray}
with the initial ratio $\zeta_0^2$.
A reasonable value of the factorization scale $\mu_f$
is smaller than, and of order $m_B$ \cite{LSW12}.
We choose $\mu_f = a \, \zeta_0  \, m_B$,
which will simplify the numerical analysis of the resummation effect in
the next section. As a consequence, the upper bound of $\tilde\zeta^2$ is replaced
by $N^2 \zeta_0^2$ in the limit $\zeta^2\to\infty$ under the requirement $\mu_f > \mu_c$,
and the scheme dependence of the $B$ meson wave function disappears.
One more advantage of the above choice of $\mu_f$ is that
the effect from resumming the rapidity logarithms does not alter the
normalization of the $B$ meson wave function for arbitrary $a$:
as $b\to 0$, which corresponds to the integration over $k_T$, and
$N\to 1$, which corresponds to the integration over $x$, the upper bound
$N^2\zeta_0^2$ is identical to the lower bound $\zeta_0^2$, and the exponential
in Eq.~(\ref{scheme solution}) becomes unity.
We stress that the normalization of the $B$ meson wave function
remains unchanged for arbitrary $\zeta^2$ as $a=1$,
which is the value of $a$ we will work
with in the next section. In this case the soft kernel in
Eq.~(\ref{improved scheme kernel}) vanishes: we have
$\tilde{K}^{(1)}(1,0,\zeta^2,\mu_c)=0$, because
the expansion of the Bessel function at small argument cancels
the logarithm $\ln (\mu_c b)$ exactly, and the integral diminishes
due to $\mu_c= \sqrt{\zeta^2}\,m_B\ge\mu_f=\zeta_0m_B$.

For given $\mu_f$, the combination of Eqs.~(\ref{scale solution}) and
(\ref{scheme solution}) gives
\begin{eqnarray}
\tilde{\Phi}_B(N,b)&=&\exp \left [
\int_{\zeta_0^2}^{N^2\zeta_0^2} { d \tilde{\zeta}^2 \over
\tilde{\zeta}^2 } {\cal K}^{(1)}(N,b,\tilde{\zeta}^2,\mu_f)
-\int^{\mu_f}_{\mu_0} \frac{d\mu}{\mu} \,
{\alpha_s(\mu) \over 2 \pi}C_F  \left ( \ln \zeta_0^2 -2 \right) \right] \nonumber \\
&& \times  \tilde{\Phi}_B(N,b,\zeta_0^2,\mu_0),\label{solu}
\end{eqnarray}
where the $\zeta^2$ dependence has disappeared in the $\zeta^2\to\infty$
limit. In this work we focus on the resummation of the double logarithms $\ln^2\zeta^2$,
and discuss its effect on the shape of the $B$ meson wave function.
For this purpose, we expand the Bessel function in Eq.~(\ref{K0}), and the
$\ln b$ term is absent. The evolution kernel then reads
\begin{eqnarray}
{\cal K}^{(1)}(N,b,\zeta^2,\mu_f)= -  {\alpha_s(\mu_c) \over 2
\pi} \, C_F \, \ln a
-\int^{\mu_f}_{\mu_c} \frac{d\mu}{\mu}  \, {\alpha_s(\mu) \over 2
\pi}C_F \, \theta(\mu_f-\mu_c).\label{k30}
\end{eqnarray}
We shall study the solution with the above simplified kernel numerically below.

\section{RESUMMATION IMPROVED WAVE FUNCTIONS}

In this section we perform the inverse Mellin transformation of
Eq.~(\ref{solu}) to obtain the $x$ dependence of the resummation
improved $B$ meson wave functions
\begin{eqnarray}
{\Phi}_B^{\pm}(x,k_T)=\int_{c-i\infty}^{c+i\infty}\frac{dN}{2\pi i} (1-x)^{-N}
\tilde{\Phi}_B^{\pm}(N,k_T),
\end{eqnarray}
where the constant $c$ is chosen such that all poles of
$\tilde{\Phi}_B^{\pm}(N,k_T)$ appear to the left of the contour in the $N$ plane.
Note that the exponential in Eq.~(\ref{solu}) involves a branch cut on the
negative real axis. We shall choose the contour which runs from minus infinity
toward the origin below the branch cut, turns around at the origin along an
infinitesimal circle, and then runs toward minus infinity above the branch cut.
With the simplified kernel in Eq.~(\ref{k30}), the Fourier transformation for
the $k_T$ dependence does not need to apply. We shall drop the RG evolution
factor, which is irrelevant to the inverse Mellin transformation, and
suppress the argument $\mu_0$. For the wave functions without the resummation,
namely, the initial conditions of the $\zeta^2$ evolution,
we take the models proposed in \cite{Kawamura:2001jm} as an example. To make transparent
the suppression mechanism near $x=0$, we first consider the case with a fixed
coupling constant. It will be explained how the exponential with the
double rapidity logarithms diminishes the $B$ meson wave function
$\Phi_B^-$ at $x=0$, which takes a finite value originally.
We then investigate the case with a running coupling constant.
Though the analysis becomes more complicated, the features of the resummation improved
$B$ meson wave functions remain.

\subsection{Resummation with fixed $\alpha_s$}

Freezing the strong coupling constant $\alpha_s$,
the solution with the resummation of the rapidity logarithms reduces to
\begin{eqnarray}
\tilde{\Phi}_B^{\pm}(N,k_T)=  \exp \left [ - {\alpha_s C_F \over  2 \, \pi}
\ln N \bigg ( \ln a^2  +  \ln N  \bigg ) \right ] \,
\tilde{\Phi}_B^{\pm}(N,k_T,\zeta_0^2),\label{fixc}
\end{eqnarray}
where the integration over $\tilde{\zeta}^2$ has been worked out.
We assume that the $B$ meson wave function
$\tilde{\Phi}_{B}^{-} (N,k_T,\zeta_0^2)$ possesses the
factorized form
\begin{eqnarray}
\Phi_{B}^{\pm}(x,k_T,\zeta_0^2) =  \phi_{B}^{\pm}(x, \zeta_0^2) \,
\phi(k_T) \,,
\end{eqnarray}
with the models in \cite{Kawamura:2001jm}
\begin{eqnarray}
\phi_{B}^{-}(x, \zeta_0^2)&=& { 2 x_0 -x \over 2 x_0^2} \, \theta(2 x_0-x),\label{bphi}\\
\phi_{B}^{+}(x, \zeta_0^2) &=& { x \over 2 x_0^2} \, \theta(2 x_0-x). \label{phipi}
\end{eqnarray}
The Mellin transformation of the above models give
\begin{eqnarray}
\tilde{\Phi}_{B}^{-}(N, k_T, \zeta_0^2)&=& { (1-2 x_0)^{N+1} + 2 x_0 N + 2 x_0-1
\over  2 x_0^2 N(N+1)} \, \phi(k_T),\\
\tilde{\Phi}_{B}^{+}(N, k_T, \zeta_0^2)&=& {1- (1-2 x_0)^{N} (1+ 2 x_0 N) \over  2 x_0^2 N(N+1)}
 \, \phi(k_T).
\end{eqnarray}

\begin{figure}[ht]
\begin{center}
\hspace{-1 cm}
\includegraphics[scale=0.5]{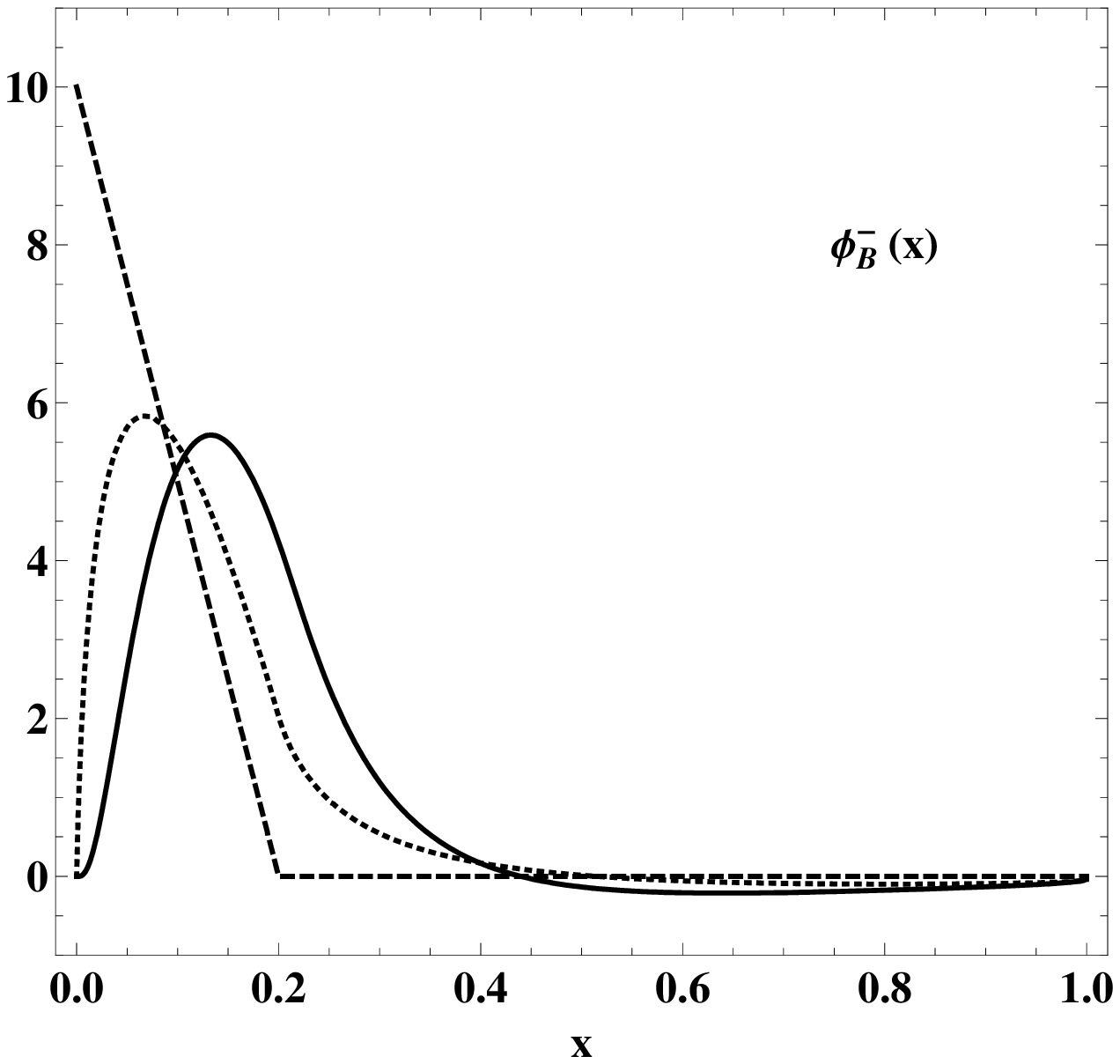}\hspace{0.5cm}
\includegraphics[scale=0.5]{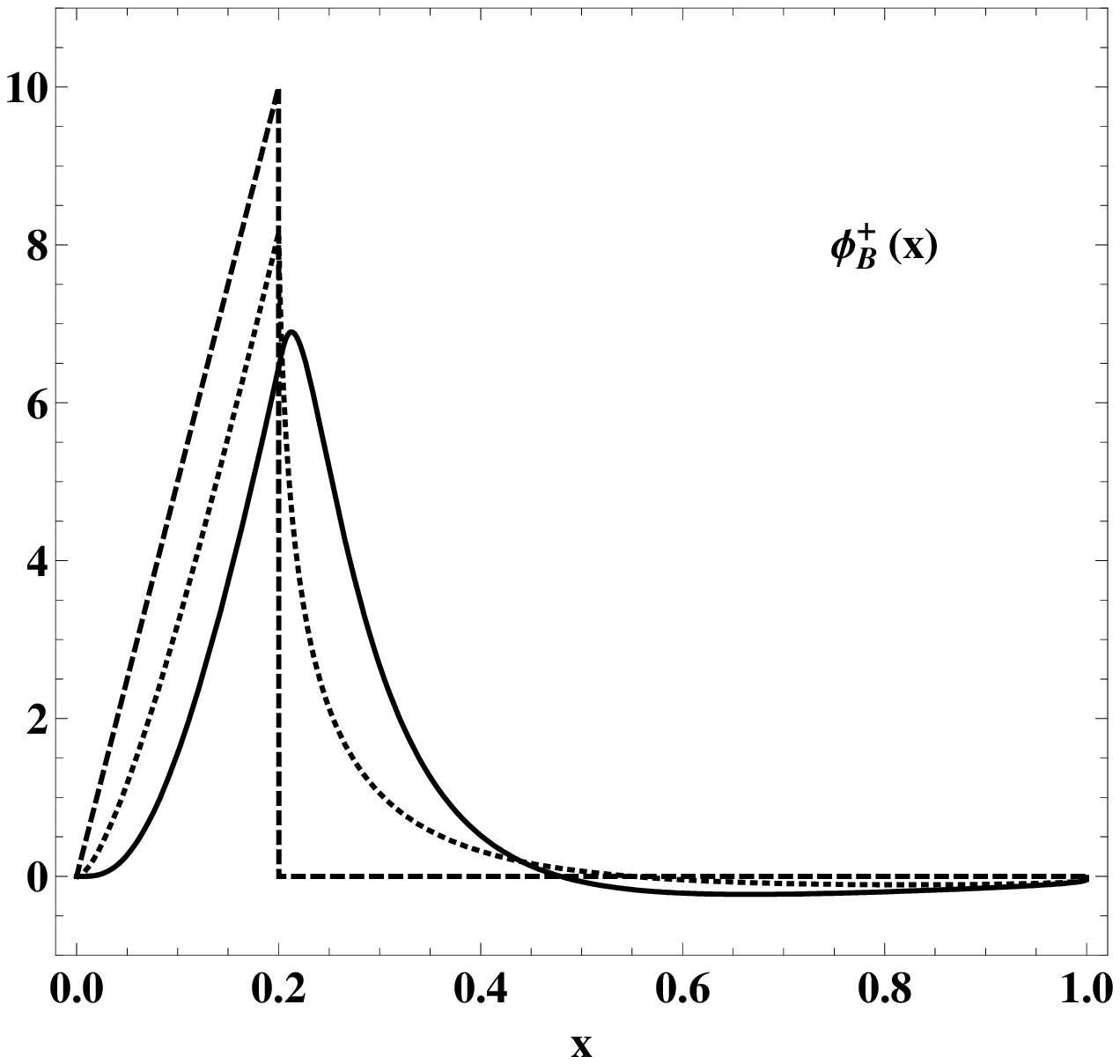}
\caption{$x$ dependence of the $B$ meson distribution amplitudes.
The dashed, dotted and solid curves correspond to the initial condition
$\phi_{B}^{\pm}(x, \zeta_0^2)$, and the resummation improved
$\phi_{B}^{\pm}(x)$ for fixed $\alpha_s=0.3$
and for running $\alpha_s$ with $\zeta_0=e/10$ and $a=1$, respectively.  }
\label{fig: shape of phiBm}
\end{center}
\end{figure}

Performing the inverse Mellin transformation, we derive the resummation
improved $B$ meson wave function
$\Phi_{B}^{-}(x, k_T)=\phi_{B}^{-}(x)\phi(k_T)$ with the distribution amplitude
\begin{eqnarray}
\phi_{B}^{-}(x) &=&
{ 1-x \over  2 x_0^2} \, \exp \left({\alpha_s C_F \over  2 \, \pi} \right)
\cos \left({\alpha_s C_F \over  2} \ln a^2  \right) \,
\theta(2 x_0-x) \nonumber \\
&& - { 1 \over  2 x_0^2} \int_{-\infty}^{+\infty} \, {d t  \over \pi} \, { (1-x)^{e^t} \over 1-e^t }
\left [ (1-2 x_0)^{1-e^t}  \theta(x-2 x_0) - 2 x_0 e^t + 2 x_0-1 \right ]   \,  \nonumber \\
&& \times \exp \left [- {\alpha_s C_F \over  2 \, \pi} \left ( t^2 +  t \, \ln a^2 - \pi^2 \right )  \right ]
\sin \left [ \alpha_s C_F (t + \ln a)  \right], \label{phim}
\end{eqnarray}
in the momentum fraction space. The first term in the above expression arises from the
$N=-1$ pole, and the second term containing an integral arises from the discontinuity
of the integrand
along the branching cut on the negative real axis. The variable change $N=\exp(t+i\pi)$
($N=\exp(t-i\pi)$) has been applied to the contour above (below) the branch cut.
The exponential with the exponent $-\ln^2 N$ in
Eq.~(\ref{fixc}) suppresses the residue of the $N=0$ pole, and diminishes
the $B$ meson wave function at $x=0$ \cite{Li02}. It is then easy to
realize that this result is not modified by variation of $a$, because of
$\ln^2N\gg \ln a^2\ln N$. The step function $\theta(x-2 x_0)$
in the square bracket is attributed to the vanishing of the inverse Mellin transformation
of $(1-2 x_0)^{N+1}/[N(N+1)]$ as $x < 2 x_0$: in this case we must close the contour
in the $N$ plane through the positive real axis, such that the semicircle at infinity
does not contribute. Hence, this contour does not enclose any poles, and the considered
inverse Mellin transformation gives a null value. The $x$ dependence of the
resummation improved $\phi_{B}^{-}(x)$ with the example set of parameters
$\alpha_s=0.3$, $\zeta_0=e/10$ and $a=1$ is displayed
in Fig.~\ref{fig: shape of phiBm}: $\phi_{B}^{-}(x)$ indeed vanishes
at $x=0$, and becomes smooth with a quick descending at large $x$,
even though the initial condition
contains a step function. It is trivial to verify that the normalization condition
$\int_0^1  d x\, \phi_{B}^{-}(x) =1$
is respected, albeit with the negative value of $\phi_{B}^{-}(x)$ at large $x$.
Moreover, we find that the resummation improved
$\phi_{B}^{-}(x)$ decreases faster than $x$ at small $x$,
$d\phi_{B}^{-}(x)/dx|_{x=0}=0$.

Following the same procedure, we derive the resummation improved wave function
$\Phi_{B}^{+}(x, k_T)=\phi_{B}^{+}(x) \phi(k_T)$ with the
distribution amplitude
\begin{eqnarray}
\phi_{B}^{+}(x) &=&
-{ 1-x \over  2 x_0^2} \, \exp \left({\alpha_s C_F \over  2 \, \pi} \right)
\cos \left({\alpha_s C_F \over  2} \ln a^2  \right) \,
\theta(2 x_0-x) \nonumber \\
&& - { 1 \over  2 x_0^2} \int_{-\infty}^{+\infty} \, {d t  \over \pi} \, { (1-x)^{e^t} \over 1-e^t }
\left [ 1-  (1-2 x_0)^{-e^t}  (1- 2 x_0 \, e^t) \, \theta(x-2 x_0)  \right ]   \,  \nonumber \\
&& \times \exp \left [- {\alpha_s C_F \over  2 \, \pi} \left ( t^2 +  t \, \ln a^2 - \pi^2 \right )  \right ]
\sin \left [ \alpha_s C_F (t + \ln a)  \right].
\end{eqnarray}
A similar reason leads to the observation that $\phi_{B}^{+}(x)$ and its first derivative
vanish at $x=0$, i.e., $\phi_{B}^{+}(0)=0$ and $d\phi_{B}^{+}(x)/dx|_{x=0}=0$,
which is not revised by variation of $a$. Namely, the resummation improved
$\phi_{B}^{+}(x)$ decreases faster than the initial condition in Eq.~(\ref{phipi})
as $x$ approaches zero. The shape of $\phi_{B}^{+}(x)$  with $\alpha_s=0.3$, $\zeta_0=e/10$ and
$a=1$ is displayed in Fig~\ref{fig: shape of phiBm}, which also becomes
smooth with a quick descending at large $x$. We have checked that the
normalization $\int_0^1 d x\, \phi_{B}^{+}(x)(x) =1$ is not
changed by the resummation effect. It is worth  mentioning that the model
of the $B$ meson wave function proposed in \cite{KLS}, which decreases like
$x^2$ at small $x$ and exhibits a Gaussian tail at large $x$, agrees with
the above features.

\subsection{Resummation with running  $\alpha_s$}

We now investigate the resummation effect with a running coupling constant $\alpha_s$.
To avoid the Landau singularity, we follow  the analytic
parametrization proposed in \cite{Solovtsov:1999in}
\begin{eqnarray}
\alpha_s(\mu)={4 \pi \over \beta_0} \left [ {1 \over \ln (\mu^2/\Lambda^2_{\rm QCD})}
- {\Lambda^2_{\rm QCD} \over \mu^2 -\Lambda^2_{\rm QCD}}   \right ],
\end{eqnarray}
to one-loop level, where $\Lambda_{\rm QCD}$ is the QCD scale, and
$\beta_0= (11 N_c - 2 N_f)/3$ is the first coefficient
of the $\beta$ function, with $N_c$ and $N_f$ being the numbers
of colors and flavors, respectively. Concentrating only on the resummation
effect, we derive the following solution from Eq.~(\ref{solu})
\begin{eqnarray}
\tilde{\Phi}_B^{\pm}(N,b) &=& \exp \left [
-{2 C_F \over \beta_0} ( A_1 \ln a  + B_1  ) \right ] \, \tilde{\Phi}_B^{\pm}(N,b,\zeta_0^2),
\end{eqnarray}
where the functions $A_1$ and  $B_1$ are
\begin{eqnarray}
A_1&=& \ln {\hat{\mu}_f \over \hat{\mu}_f  - \ln N}
+ \ln {\mu_f^2 - N^2 \Lambda_{\rm QCD}^2 \over  \mu_f^2 - \Lambda_{\rm QCD}^2}, \nonumber \\
B_1&=&  {1 \over 2}\ln^2 N + (\hat{\mu}_f - \ln N) \, \ln (\hat{\mu}_f- \ln N)
- \hat{\mu}_f \ln \hat{\mu}_f  + \ln N \left ( \ln {\mu_f \, \Lambda_{\rm QCD} \over  \mu_f^2 - \Lambda_{\rm QCD}^2}
+ \ln \hat{\mu}_f + 1 \right ) \nonumber  \\
&& - {\rm Li}_2\left (- { \mu_f \over \Lambda_{\rm QCD} } \right )
+ {\rm Li}_2\left (- { \mu_f \over N \Lambda_{\rm QCD} } \right )
-{\rm Li}_2\left (- { N \, \Lambda_{\rm QCD} \over \mu_f } \right )
+ {\rm Li}_2\left (- { \Lambda_{\rm QCD} \over \mu_f } \right ) ,
\end{eqnarray}
with $\hat{\mu}_f= \ln (\mu_f/\Lambda_{\rm QCD})$.

Employing the models in Eqs.~(\ref{bphi}) and
(\ref{phipi}) for the initial conditions,
and performing the inverse Mellin transformation of $\tilde{\Phi}_B^{\pm}(N,k_T)$,
we finally arrive at the distribution amplitudes
\begin{eqnarray}
\phi_{B}^{-}(x) &=& {1- x \over 2 x_0^2} \,
\exp \left[  - {2 C_F \over \beta_0}  (A_2 \ln a +B_2)  \right ]
\, \cos \left[  - {2 C_F \over \beta_0}  (A_3 \ln a +B_3)  \right ]\, \theta(2 x_0-x)
\nonumber \\
&& -{ 1 \over  2 x_0^2} \int_{-\infty}^{+\infty} \, {d t  \over \pi} \, { (1-x)^{e^t} \over 1-e^t }
\left [ (1-2 x_0)^{1-e^t}  \theta(x-2 x_0) - 2 x_0 e^t + 2 x_0-1 \right ]   \,
\nonumber \\
&& \times \exp \left [  - {2 C_F \over \beta_0}  (A_4 \ln a +B_4)  \right ]
\, \sin \left [  - {2 C_F \over \beta_0}  (A_5 \ln a +B_5)  \right ],\label{ab1}\\
\phi_{B}^{+}(x) &=& - {1- x \over 2 x_0^2} \,
\exp \left [  - {2 C_F \over \beta_0}  (A_2 \ln a +B_2)  \right ]
\, \cos \left [  - {2 C_F \over \beta_0}  (A_3 \ln a +B_3)  \right ]\, \theta(2 x_0-x)
\nonumber \\
&& -{ 1 \over  2 x_0^2} \int_{-\infty}^{+\infty} \, {d t  \over \pi} \, { (1-x)^{e^t} \over 1-e^t }
\left [ 1-  (1-2 x_0)^{-e^t}  (1- 2 x_0 \, e^t) \, \theta(x-2 x_0)  \right ]   \,
\nonumber \\
&& \times\exp \left [  - {2 C_F \over \beta_0}  (A_4 \ln a +B_4)  \right ]
\, \sin \left [  - {2 C_F \over \beta_0}  (A_5 \ln a +B_5)  \right ]\label{ab2} ,
\end{eqnarray}
where the functions $A_i$ and $B_i$ ($i=2$-5) are collected in Appendix A.
Their $x$ dependencies are also shown in Fig.~\ref{fig: shape of phiBm}, which
do not differ much from the curves for the fixed $\alpha_s$, but with the peak
positions shifting toward large $x$ a bit. The normalization condition
and the suppression at $x=0$, $\phi_{B}^{\pm}(0) =0$ and $d\phi_{B}^{\pm}(x)/dx|_{x=0}=0$,
are maintained. We mention that the ``Wandzura-Wilczek relations" \cite{Wandzura:1977qf}
of the two $B$ meson wave functions do not hold anymore after including the resummation
of the rapidity logarithms. This observation is not unexpected, since it is not clear
that such relations sustain under radiative corrections to all orders
\cite{Beneke:2000wa,Bell:2008er}.

\begin{figure}[ht]
\begin{center}
\hspace{-1 cm}
\includegraphics[scale=0.6]{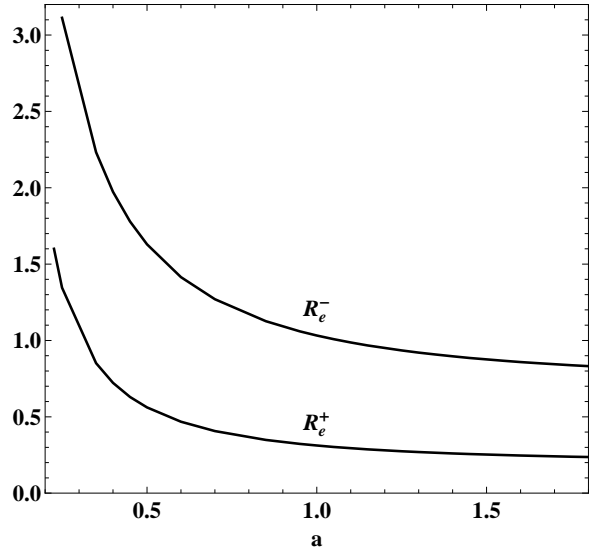}
\caption{$a$ dependence of the ratios $R^{\pm}_{e}$ for $\zeta_0=e/10$.}
\label{fig: resummation effect}
\end{center}
\end{figure}

To illustrate the $a$ dependence of the resummation effect,
we consider the ratios of the resummation improved $B$ meson distribution amplitudes
over the initial conditions for $x=0.1$,
\begin{eqnarray}
R^{\pm}_{e}=\frac{\phi_{B}^{\pm}(0.1)}{\phi_{B}^{\pm}(0.1,\zeta_0^2)}.
\end{eqnarray}
These ratios for $\zeta_0=e/10$ are displayed
in Fig.~\ref{fig: resummation effect}, which decrease with $a$ and become
stable as $a>1$.  It implies that the peaks of $\phi_{B}^{\pm}(x)$
will be pushed toward large $x$, as $a$ increases. However,
as long as $a$ remains of order unity, namely, away from zero, the shape change
will not be significant. In this case
the characteristic scale $\mu_c\sim \, O(\sqrt{\zeta^2}\,m_B/N)$
removes the large logarithm in the initial condition $K^{(1)}(N,b,\zeta^2,\mu_c)$,
and the kernel governing the resummation effect
roughly stays the same.

\begin{figure}[ht]
\begin{center}
\hspace{-1 cm}
\includegraphics[scale=0.6]{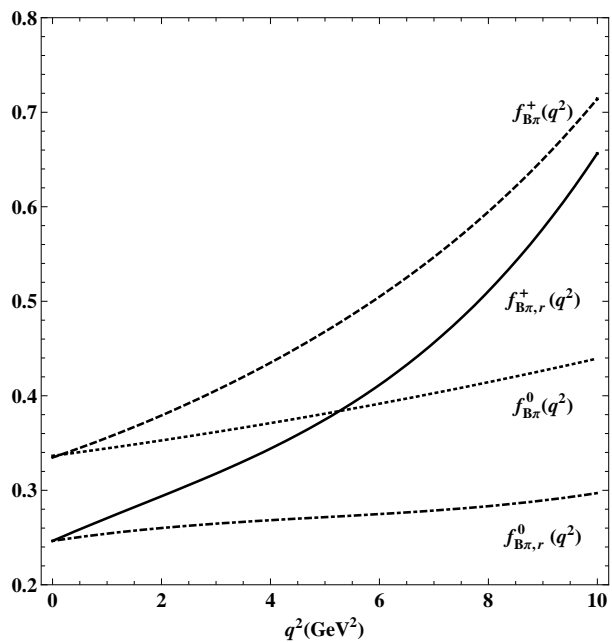}
\caption{$q^2$ dependence of the $B \to \pi$ form factors $f_{B \to \pi}^{+}(q^2)$
and $f_{B \to \pi}^{0}(q^2)$ in the $k_T$ factorization with
the resummation effect (solid and dash-dotted lines, respectively) and without the
resummation effect (dashed and dotted lines, respectively). }
\label{fig: shape of B to pi form factors}
\end{center}
\end{figure}

A hard kernel $H$, defined as the difference between the QCD quark diagrams
and the effective diagrams, also contains the rapidity logarithms. In principle,
these logarithms should be resummed by solving the evolution equation
\begin{eqnarray}
\zeta^2\frac{d}{d\zeta^2}\tilde H(N,b,\zeta^2,\mu_f)=-
\tilde K^{(1)}_H(N,b,\zeta^2,\mu_f) \, \tilde H(N,b,\zeta^2,\mu_f),
\label{hevo}
\end{eqnarray}
which arises from the $\zeta^2$ independence of the QCD quark diagrams. 
After treating $\ln\zeta^2$ in both
the $B$ meson wave functions and the hard kernel, the scheme dependence
disappears completely. Strictly speaking, the $\zeta^2$ dependence
in the pion wave functions should be treated too. The above subjects will be studied
in a separate work. Here we investigate the resummation effect
on the $B \to \pi$ transition form factors involved in semileptonic
decays \cite{LSW12}. For this purpose, it is legitimate to simply
convolute the LO hard kernel with the $B$ meson wave functions before and after the
resummation. We take this opportunity to correct a typo in Eq. (35)  of \cite{LSW12},
where the $+ (1/2) \ln (\zeta_1^2/ m_B^2) $ term  should be replaced by
$- (1/ 2) \ln (\zeta_1^2/ m_B^2) $. Correspondingly,  the term
$- (1/ 2) \ln (m_B^2/ \zeta_1^2) \left [ 3 \ln (m_B^2/ \zeta_1^2) +2 \right ] $
in Eqs. (54) and (56) of \cite{LSW12} needs to be changed into
$-(1/ 2) \ln (m_B^2/ \zeta_1^2) \left [  \ln (m_B^2 /\zeta_1^2) +2 \right ] $,
and the term $+ (1/ 2) \ln (m_B^2/ \zeta_1^2) \left[ 3 \ln (m_B^2/ \zeta_1^2) +2 \right ] $
involved in the coefficient $c_1$ in Eq. (61) should be
$+ (1/ 2) \ln(m_B^2/\zeta_1^2) \left[\ln (m_B^2/ \zeta_1^2) +2 \right ] $.
The $B \to \pi$ form factors $f_{B \to \pi}^{+}(q^2)$
and $f_{B \to \pi}^{0}(q^2)$ on the lepton-pair invariant mass $q^2$ in the
$k_T$ factorizaton approach are presented in
Fig.~\ref{fig: shape of B to pi form factors}, where the RG evolution effect
in Eq.~(\ref{solu}) has been included. It is observed that the resummation effect
decreases both form factors by about 25\% at $q^2=0$, which is attributed to the suppression
of the end-point behavior of the $B$ meson wave functions.

\section{Conclusion}

In this paper we have constructed the evolution equation for the $B$ meson
wave functions in the $k_T$ factorization theorem, whose solution sums the rapidity
logarithms from the light-cone singularities. Since the $B$ meson wave functions
are defined with the rescaled $b$-quark field, the ultraviolet behavior of the
evolution kernel differs from that in conventional resummations.
It has been shown that the RG evolution
in the factorization scale must be taken into account simultaneously,
in order to have a well-defined evolution kernel. The commutativity
between the scheme and RG evolutions demands the RG invariance of
the evolution kernel, which is the key step to perform the resummation.
It is interesting to notice that this new resummation formalism is similar to
the joint resummation, which combines the threshold and $k_T$ resummations.
We have demonstrated that the resummation effect respects the normalization
of the $B$ meson wave functions, and strengths their convergent behavior near
the end point of the spectator momentum. These features cannot be obtained
in the collinear factorization theorem, in which the $B$ meson distribution amplitude
loses the normalizability under the RG evolution.
It has been found that the resummation improved $B$ meson wave functions
decrease the $B\to\pi$ transition form factors in the $k_T$ factorization theorem,
which is attributed to the suppression at the end point.

\section*{Acknowledgement}

We would like to thank Thorsten Feldmann and Tao Huang for helpful discussions.
This work was supported in part by the National Science Council of
R.O.C. under Grant No. NSC-101-2112-M-001-006-MY3, by the National
Center for Theoretical Sciences of R.O.C., by National Science
Foundation of China under Grant No. 11005100, by the German
research foundation DFG under contract MA1187/10-1, and by the
German Ministry of Research (BMBF) under contract 05H09PSF.

\appendix

\section{Explicit expressions of the functions $A_i$ and $B_i$}
\label{functions A and B}

In this appendix we collect the explicit expressions of the functions $A_i$ and $B_i$
appearing in Eqs.~(\ref{ab1}) and (\ref{ab2}):
\begin{eqnarray}
A_2 = \ln \hat{\mu}_f - {1 \over 2} \ln \left (\hat{\mu}_f^2 + \pi^2 \right )\,,  & \qquad & A_3 = \phi_1 \,,  \nonumber \\
A_4 = {1 \over 2} \ln \left [{\hat{\mu}_f^2  \over  (\hat{\mu}_f-t)^2+\pi^2 } \right]
+ \ln \left| { \mu_f^2 -e^{2 t} \Lambda_{\rm QCD}^2 \over \mu_f^2 - \Lambda_{\rm QCD}^2 } \right|\,,
& \qquad &  A_5 =  - \phi_2 - \pi \theta(t- \hat{\mu}_f) \,. \hspace{0.5 cm}
\end{eqnarray}
\begin{eqnarray}
B_2 &=& -{\pi^2 \over 2} + {1 \over 2} \hat{\mu}_f  \ln \left (\hat{\mu}_f^2 + \pi^2 \right )
- \pi \, \phi_1 - \hat{\mu}_f \ln \hat{\mu}_f  -  {\rm Li}_2\left (- { \mu_f \over \Lambda_{\rm QCD} } \right )
\nonumber \\
&& +{\rm Re} \left [  {\rm Li}_2\left ({ \mu_f \over \Lambda_{\rm QCD} } \right ) \right ]
- {\rm Li}_2\left (- {  \Lambda_{\rm QCD} \over \mu_f } \right )
+ {\rm Li}_2\left ( {  \Lambda_{\rm QCD} \over \mu_f } \right ) \,, \nonumber \\
B_3 &=& -{\pi^2 \over 2} \ln \left (\hat{\mu}_f^2 + \pi^2 \right )
+ \pi \left ( \ln {\mu_f^2 \over  \mu_f^2 - \Lambda_{\rm QCD}^2}
+ \ln \hat{\mu}_f + 1 \right )- \phi_1  \, \hat{\mu}_f\,, \nonumber \\
B_4 &=& {t^2 - \pi^2 \over 2} + {\hat{\mu}_f -t \over 2}  \ln \left [(\hat{\mu}_f -t)^2 + \pi^2 \right]
- \pi \phi_2 - \hat{\mu}_f \ln \hat{\mu}_f  \nonumber \\
&& + t \left ( \ln {\mu_f \Lambda_{\rm QCD}  \over  \mu_f^2 - \Lambda_{\rm QCD}^2}
+ \ln \hat{\mu}_f + 1 \right )
-  {\rm Li}_2\left (- { \mu_f \over \Lambda_{\rm QCD} } \right )
\nonumber \\
&& +{\rm Re} \left [  {\rm Li}_2\left ({ \mu_f \over e^t \Lambda_{\rm QCD} } \right ) \right ]
- {\rm Li}_2\left (- {  e^t \Lambda_{\rm QCD} \over \mu_f } \right )
+ {\rm Li}_2\left ( {  \Lambda_{\rm QCD} \over \mu_f } \right ) \,, \nonumber \\
B_5 &=& - \pi t  + {\pi \over 2} \ln \left[ (\hat{\mu}_f -t)^2 + \pi^2 \right]
   - \pi \left ( \ln {\mu_f \Lambda_{\rm QCD}  \over  \mu_f^2 - \Lambda_{\rm QCD}^2}
+ \ln \hat{\mu}_f + 1 \right ) \nonumber \\
&& + \phi_2 (\hat{\mu}_f -t)  - \pi (\hat{\mu}_f -t) \theta(\hat{\mu}_f -t) \,,
\end{eqnarray}
with
\begin{eqnarray}
\phi_1 = {\rm arctan} {\pi \over \hat{\mu}_f }\,, \qquad
\phi_2 ={\rm arctan} {\pi \over \hat{\mu}_f  -t} \, \theta(\hat{\mu}_f -t)
+ \left [\pi +{\rm arctan} {\pi \over \hat{\mu}_f  -t} \right ] \, \theta(t-\hat{\mu}_f) \,.
\nonumber
\end{eqnarray}


\begin{thebibliography}{99}



\bibitem{BL} G.P. Lepage and S.J. Brodsky, Phys. Lett. B {\bf 87},
359 (1979); Phys. Rev. D {\bf 22}, 2157 (1980).



\bibitem{ER} A.V. Efremov and A.V. Radyushkin, Phys. Lett. B {\bf 94},
245 (1980).




\bibitem{CZS} V.L. Chernyak, A.R. Zhitnitsky, and V.G. Serbo,
JETP Lett. {\bf 26}, 594 (1977).




\bibitem{CZ} V.L. Chernyak and A.R. Zhitnitsky,
Sov. J. Nucl. Phys. {\bf 31}, 544 (1980); Phys. Rep. {\bf 112}, 173
(1984).



\bibitem{L1} H-n. Li, Phys. Rev. D {\bf 64}, 014019 (2001); M. Nagashima
and H-n. Li, Eur.\ Phys.\ J.\ C {\bf 40} (2005) 395.




\bibitem{GN} A.G. Grozin and M. Neubert, Phys. Rev. D {\bf 55},
272 (1997).





\bibitem{PB1} V.M. Braun and I.E. Filyanov, Z. Phys. C {\bf 48}, 239
(1990); P. Ball, JHEP {\bf 9901}, 010 (1999).





\bibitem{Braun:1990iv} P. Ball, V.M. Braun, Y. Koike and K. Tanaka,
Nucl. Phys. {\bf B529}, 323 (1998).





\bibitem{Kawamura:2001jm} H. Kawamura, J. Kodaira, C.F. Qiao, and K. Tanaka,
Phys. Lett. B {\bf 523}, 111 (2001); Erratum-ibid. 536, 344 (2002);
Mod. Phys. Lett. A {\bf 18}, 799 (2003).



\bibitem{ML99} H.-n. Li and B. Melic,
Eur. Phys. J. C {\bf 11}, 695 (1999).

\bibitem{CL05} Y.Y. Charng and H.-n. Li
Phys. Rev. D {\bf 72}, 014003 (2005).




\bibitem{Neu03} B.O. Lange and M. Neubert, Phys. Rev. Lett. {\bf 91},
102001 (2003).



\bibitem{BIK} V.M. Braun, D.Yu. Ivanov, G.P. Korchemsky,
Phys. Rev. D {\bf 69}  (2004) 034014.


















\bibitem{BBNS} M. Beneke, G. Buchalla, M. Neubert, and C.T. Sachrajda,
Phys. Rev. Lett. {\bf 83}, 1914 (1999);
Nucl. Phys. {\bf B591}, 313 (2000); Nucl. Phys. {\bf B606}, 245 (2001).


\bibitem{PK} P. Ball and E. Kou, JHEP {\bf 0304}, 029 (2003).



\bibitem{Li:2004ja}
  H.~-n.~Li and H.~-S.~Liao,
  Phys.\ Rev.\ D {\bf 70} (2004) 074030.





\bibitem{SHB} A. Szczepaniak, E.M. Henley, and S. Brodsky,
Phys. Lett. B {\bf 243}, 287 (1990).








\bibitem{CCH} S. Catani, M. Ciafaloni and F. Hautmann, Phys. Lett.
B {\bf 242}, 97 (1990); Nucl. Phys. B {\bf 366}, 135 (1991).




\bibitem{CE} J.C. Collins and R.K. Ellis, Nucl. Phys. B {\bf 360}, 3 (1991).




\bibitem{LRS} E.M. Levin, M.G. Ryskin, Yu.M. Shabelskii,
and A.G. Shuvaev, Sov. J. Nucl. Phys. {\bf 53}, 657 (1991).




\bibitem{BS} J. Botts and G. Sterman, Nucl. Phys. B {\bf 325}, 62 (1989).



\bibitem{LS} H.-n. Li and G. Sterman, Nucl. Phys. B {\bf 381}, 129 (1992).



\bibitem{NL2} M. Nagashima and H-n. Li, Phys. Rev. D {\bf 67},
034001 (2003).




\bibitem{Monr} H-n. Li, hep-ph/0304217.




\bibitem{HS} T. Huang and Q.-X. Shen, Z. Phys. C {\bf 50}, 139 (1991);
J.P. Ralston and B. Pire, Phys. Rev. Lett. {\bf 65}, 2343 (1990);
R. Jakob and P. Kroll, Phys. Lett. B {\bf 315}, 463 (1993); B {\bf
319}, 545 (1993)(E).



\bibitem{LY1} H-n. Li and H.L. Yu, Phys. Rev. Lett. {\bf 74}, 4388 (1995);
Phys. Lett. B {\bf 353}, 301 (1995); Phys. Rev. D {\bf 53}, 2480 (1996);
H-n. Li, Chin. J. Phys. {\bf 34}, 1047 (1996).




\bibitem{CL} C.H. Chang and H-n. Li, Phys. Rev. D {\bf 55}, 5577
(1997); T.W. Yeh and H-n. Li, Phys. Rev. D {\bf 56}, 1615 (1997).




\bibitem{KLS} Y.Y. Keum, H-n. Li, and A.I. Sanda,
Phys. Lett. B {\bf 504}, 6 (2001); Phys. Rev. D {\bf 63}, 054008 (2001);
Y.Y. Keum and H-n. Li, Phys. Rev. {\bf D63}, 074006 (2001).




\bibitem{LUY} C. D. L\"{u}, K. Ukai, and M. Z. Yang, Phys. Rev. D {\bf 63},
074009 (2001).




\bibitem{TLS} T. Kurimoto, H-n. Li, and A.I. Sanda,
Phys. Rev D {\bf 65}, 014007 (2002); Phys. Rev. D {\bf 67}, 054028
(2003).




\bibitem{KS02} Y.Y. Keum, H-n. Li, and A.I. Sanda,
AIP Conf. Proc. {\bf 618}, 229 (2002);
Y.Y. Keum and A.I. Sanda, Phys. Rev. D {\bf 67},
054009 (2003).

\bibitem{Li07} H.-n. Li, ECONFC070512:011 (2007)
[arXiv:0707.1294 [hep-ph]].




\bibitem{Ali:2007ff}
  X.~-G.~He, T.~Li, X.~-Q.~Li and Y.~-M.~Wang,
  Phys.\ Rev.\ D {\bf 75} (2007) 034011;
  P.~Guo, H.~-W.~Ke, X.~-Q.~Li, C.~-D.~Lu and Y.~-M.~Wang,
  Phys.\ Rev.\ D {\bf 75} (2007) 054017;
  A.~Ali, G.~Kramer, Y.~Li, C.~-D.~Lu, Y.~-L.~Shen, W.~Wang and Y.~-M.~Wang,
  Phys.\ Rev.\ D {\bf 76} (2007) 074018;
  C.~-D.~Lu, Y.~-M.~Wang, H.~Zou, A.~Ali and G.~Kramer,
  Phys.\ Rev.\ D {\bf 80} (2009) 034011.






\bibitem{CS91}
J. C. Collins and D. E. Soper,
Nucl. Phys. {\bf B193}, 381 (1981); J. C. Collins,
Adv. Ser. Direct. High Energy Phys. 5, 573 (1989).



\bibitem{Co03} J.C. Collins, Acta. Phys. Polon. B {\bf 34}, 3103 (2003).







\bibitem{LSW12} H.-n. Li, Y.-L. Shen, and Y.-M. Wang,
Phys. Rev. D {\bf 85}, 074004 (2012).


\bibitem{LSW11} H.-n. Li, Y.-L. Shen, Y.-M. Wang, and H. Zou,
Phys. Rev. D {\bf 83}, 054029 (2011).



\bibitem{Co92} J.C. Collins and F. Tkachov, Phys. Lett. B {\bf 294}, 403 (1992);
J.C. Collins, PoS LC2008, 028 (2008) [arXiv:0808.2665 [hep-ph]].

\bibitem{Chiu:2012ir}
  J.~-y.~Chiu, A.~Jain, D.~Neill and I.~Z.~Rothstein,
  Phys.Rev.Lett. {\bf 108}, 151601 (2012); JHEP {\bf 1205} (2012) 084.

\bibitem{FZ12} S. Fleming and O. Zhang, arXiv:1210.1508 [hep-ph].


\bibitem{Altinoluk:2012fb}
  T.~Altinoluk, B.~Pire, L.~Szymanowski and S.~Wallon,
  arXiv:1206.3115 [hep-ph];
  T.~Altinoluk, B.~Pire, L.~Szymanowski and S.~Wallon,
  JHEP {\bf 1210} (2012) 049
  [arXiv:1207.4609 [hep-ph]].


\bibitem{CS08} I.O. Cherednikov and N.G. Stefanis,
Nucl. Phys. {\bf B802}, 146 (2008).

\bibitem{Li:1996gi} H.-n. Li, Phys. Rev. D {\bf 55},
105 (1997)











\bibitem{Li99} H.-n. Li, Phys. Lett. B {\bf 454}, 328 (1999).




\bibitem{LSV00} E. Laenen, G. Sterman, and W. Vogelsang,
Phys. Rev. Lett. {\bf 84}, 4296 (2000).



\bibitem{DFK11} J. Debove, B. Fuks, and M. Klasen,
Nucl. Phys. {\bf B849}, 64 (2011).



\bibitem{Feng:2009rp} J.~P. Ma and Q. Wang, Phys. Lett. B {\bf 613}, 39 (2005);
  F.~Feng, J.~P.~Ma and Q.~Wang,
  arXiv:0901.2965 [hep-ph].




\bibitem{Li02} H.-n. Li,
Phys. Rev. D {\bf 66}, 094010 (2002).




\bibitem{Solovtsov:1999in}
  I.~L.~Solovtsov and D.~V.~Shirkov,
  Theor.\ Math.\ Phys.\  {\bf 120} (1999) 1220
   [Teor.\ Mat.\ Fiz.\  {\bf 120} (1999) 482].





\bibitem{Wandzura:1977qf}
  S.~Wandzura and F.~Wilczek,
  Phys.\ Lett.\ B {\bf 72} (1977) 195.




\bibitem{Beneke:2000wa}
  M.~Beneke and T.~Feldmann,
  Nucl.\ Phys.\ B {\bf 592} (2001) 3.




\bibitem{Bell:2008er}
  G.~Bell and T.~Feldmann,
  JHEP {\bf 0804} (2008) 061.









\end{thebibliography}
\end{document}